\input harvmac

\Title{\vbox{\rightline{EFI-98-13}\rightline{hep-th/9804111}}}
{\vbox{\centerline{Matrix Models of AdS Gravity}}}
\bigskip

\baselineskip=12pt
\centerline{Emil J. Martinec} 
\bigskip
\centerline{\sl Enrico Fermi Inst. and Dept. of Physics}
\centerline{\sl University of Chicago}
\centerline{\sl 5640 S. Ellis Ave., Chicago, IL 60637, USA}

\baselineskip=16pt
 
\vskip 2cm
\noindent

We explore the connection between anti-deSitter supergravity
and gauge theory, in the context of bound states of many D1 and D5 branes.
The near-horizon $AdS_3\times S^3$ supergravity
describes the identity sector of the conformal field theory
produced by the brane dynamics.
A variant of anomaly inflow (for the 2d conformal anomaly) is involved.
Dynamical matter fields on $AdS_3$
couple to the chiral ring and its
descendant fields on the branes.  
We propose a map between boundary conformal field
theory and bulk supergravity/matter dynamics, which is
strongly reminiscent of matrix models of 2d gravity.

\Date{4/98}

%
%
\def\journal#1&#2(#3){\unskip, \sl #1\ \bf #2 \rm(19#3) }
\def\andjournal#1&#2(#3){\sl #1~\bf #2 \rm (19#3) }

\def\eg{{\it e.g.}}
\def\cf{{\it c.f.}}

\def\etc{{\it etc.}}

\def\sst{\scriptscriptstyle}

\def\frac#1#2{{#1\over#2}}
\def\coeff#1#2{{\textstyle{#1\over #2}}}
\def\half{\frac12}
\def\hf{{\textstyle\half}}

\def\d{\partial}

\def\inbar{\,\vrule height1.5ex width.4pt depth0pt}
\def\IC{\relax\hbox{$\inbar\kern-.3em{\rm C}$}}
\def\IR{\relax{\rm I\kern-.18em R}}
\def\IP{\relax{\rm I\kern-.18em P}}
\def\Z{{\bf Z}}

%
%

\def\npb#1#2#3{Nucl. Phys. {\bf B#1} (#2) #3}

\def\plb#1#2#3{Phys. Lett. {\bf #1B} (#2) #3}
\def\prl#1#2#3{Phys. Rev. Lett. {\bf #1} (#2) #3}

\def\prd#1#2#3{Phys. Rev. {\bf D#1} (#2) #3}

\def\cmp#1#2#3{Comm. Math. Phys. {\bf #1} (#2) #3}
\def\cqg#1#2#3{Class. Quant. Grav. {\bf #1} (#2) #3}
\def\mpl#1#2#3{Mod. Phys. Lett. {\bf #1} (#2) #3}

\catcode`\@=11
\def\slash#1{\mathord{\mathpalette\c@ncel{#1}}}
\overfullrule=0pt
\def\AA{{\cal A}}

\def\DD{{\cal D}}
\def\EE{{\cal E}}

\def\GG{{\cal G}}

\def\MM{{\cal M}}
\def\NN{{\cal N}}
\def\OO{{\cal O}}

\def\RR{{\cal R}}

\def\VV{{\cal V}}

\def\vareps{\varepsilon}
\def\underrel#1\over#2{\mathrel{\mathop{\kern\z@#1}\limits_{#2}}}

\catcode`\@=12


%

\def\det{{\rm det}}

\def\det{{\rm det}}
\def\exp{{\rm exp}}
\def\sh{{\rm sh}}
\def\ch{{\rm ch}}

\def\bh{{\sst BH}}

\def\lpl{\ell_{{\rm pl}}}
\def\lstr{\ell_{{\rm str}}}
\def\aleff{{\alpha'_{\rm eff}}}

\def\gstr{g_{\sst\rm str}}
\def\ads{{${\it AdS}_3$}}
\def\X{{\bf X}}
\def\lz{{L_0}}
\def\lzb{{{\bar L}_0}}
\def\atil{{\tilde A}}

\def\abar{{\bar A}}

\def\sltwo{{$SL(2,R)$}}
\def\A{{\sst\rm(A)}}

\def\M{{\bf M}}
\def\aa{{\tt a}}
\def\ee{{\EE}}
\def\oo{{\Omega}}
\def\rr{{\bf r}}
\def\IH{\relax{\rm I\kern-.18em H}}
%
%

\newsec{Introduction}

The duality between open and closed string processes is an old
story \ref\opcl{E. Cremmer and J. Scherk, \npb{50}{1972}{222}.}; the
string diagrammatic expansion involving Riemann surfaces with
boundary contains boundaries of the moduli space involving
factorization on open as well as closed string states.
Thus there has been persistent speculation (see
\ref\induced{E. Witten, \npb{276}{1986}{291}.} 
for but one example)
that closed string dynamics (gravity) is somehow induced
from open string dynamics (gauge theory).  The most recent
examples are matrix theory 
\ref\bfss{T. Banks, W. Fischler, S.H. Shenker and L. Susskind,
hep-th/9610043; \prd {55}{1997}{5112}.}, 
where the leading
long-range gravitational interaction arises as an effect
of open string quantum fluctuations,
and Maldacena's proposal 
\ref\malda{J. Maldacena, hep-th/9711200.}
that anti-deSitter supergravity is in some sense `dual'
to supersymmetric gauge theory.  
In the latter case, the near-horizon
geometry of M2, M5, D3, and D1+D5 branes (in the
large $N$ limit at large, fixed $\gstr N$) 
is $AdS_p\times S^d$
for appropriate values of $p$ and $d$.  This limit holds fixed
the radii of the anti-deSitter space and the sphere 
while the Planck length $\lpl\rightarrow 0$,
so that gravity is classical.
The corresponding brane theories are conformally invariant in the
limit; their superconformal symmetries match the isometries of
$AdS_p\times S^d$.
Thus far, Maldacena's conjecture has been used to make a
series of predictions about gauge theory in the limit $N\rightarrow\infty$,
$\gstr N$ large, which are difficult to test via a corresponding
calculation in the gauge theory.

The issue at hand is whether gravity encodes 
gauge theory correlation functions, or is merely coupled to 
gauge dynamics. 
We are by now accustomed
to the idea that short-distance dynamics in one is related
to long-distance dynamics in the other (\cf\
\ref\uvir{A. Sen, hep-th/9605150, \npb{475}{1996}{562};
T. Banks, M.R. Douglas, N. Seiberg, hep-th/9605199, \plb{387}{1996}{278};
N. Seiberg, hep-th/9606017, \plb{384}{1996}{81}.}).
In the D3 brane case, it is expected 
\nref\gkp{S.S. Gubser, I.R. Klebanov, and A.M. Polyakov, hep-th/9802109.}%
\nref\wittenads{E. Witten, hep-th/9802150; hep-th/9803131.}%
\refs{\malda,\gkp,\wittenads}
that operators coupling
to non-BPS excitations of the gauge theory have scaling dimensions
that are of order $\ell/\lstr$, where $\ell$ is the
radius of the anti-deSitter space; then to discern them would involve
the stringy resonances in closed string theory.
This is what one would expect on the basis of worldsheet 
(channel) duality --
open string dynamics is only obtained upon
summing over an infinite tower of resonances
in the closed string sector. 
Truncation of the dynamics
to the lowest states takes an average over microstates of
the gauge theory.
Thus, the information encoded by classical supergravity should
be a subset of the information in the gauge dynamics.
The massless closed strings couple to the energy-momentum tensor, \etc,
of open string theory.  One therefore expects that the
classical spacetime action of the massless level of
closed string theory is the generating function 
of the insertion of currents 
and other `geometrical' operators in
the planar limit of open string field theory.


The present work analyzes a situation
where both sides of the brane/supergravity dynamics
are under some level of control, in order to study
the conjectured correspondence of 
\refs{\malda,\gkp,\wittenads}.  We will work with
the D1-D5 system in IIB string theory, since on the one hand, 
the low-energy supergravity is topological in the infrared,
and coupled to an assortment of matter fields arising from
Kaluza-Klein reduction; and on the other hand, the IR dynamics on
the branes is a 1+1 dimensional $\NN=(4,4)$ sigma model
whose properties have been extensively investigated.
For studies of this system, see for example
\nref\dodf{M.R. Douglas, hep-th/9604198;
R. Dijkgraaf, E. Verlinde, and H. Verlinde,
hep-th/9704018, \npb{506}{1997}{121};
E. Witten, hep-th/9707093;
S.F. Hassan and S. Wadia, hep-th/9703163,
\plb{402}{1997}{43}; hep-th/9712213.}%
\nref\maldathesis{J. Maldacena, Princeton Ph.D. thesis, hep-th/9607235;
see also hep-th/9705078, Nucl.Phys.Proc.Suppl. {\bf 61A} (1998) 111-123.}%
\refs{\dodf,\maldathesis}
(and 
\ref\BHreview{G. Horowitz, contribution to the Symposium
on Black Holes and Relativistic Stars
(dedicated to the memory of S. Chandrasekhar),
Chicago, IL, 14-15 Dec 1996, hep-th/9704072; A. Peet, hep-th/9712253.}
for general reviews on black holes in string theory). 
We will discuss some aspects of the IR
conformal field theory on the branes in section 2,
and of $AdS_3\times S^3$ supergravity in section 3.

In this system, there is a precise correspondence between
the {\it identity sector} of the IR conformal field theory 
on the branes and the low-energy limit
of supergravity on $AdS_3\times S^3$,
which we develop in section 4.
The core ideas are the relation of \ads\ supergravity
as a Chern-Simons theory 
\nref\hipt{A. Achucarro and P.K. Townsend, \plb{180}{1986}{89};
P.S. Howe, J.M. Izquierdo, G. Papadopoulos, and
P.K. Townsend, hep-th/9505032.}%
\nref\witcsgrav{E. Witten, \npb{311}{1988}{46}; \npb{323}{1989}{113}.}%
\refs{\hipt,\witcsgrav} 
on the group $SU(1,1|2)\times SU(1,1|2)$
to the corresponding WZW theory on the boundary
\ref\wzw{E. Witten, \cmp{121}{1989}{351};
G. Moore and N. Seiberg, \plb{220}{1989}{422};
S. Elitzur, G. Moore, A. Schwimmer, and N. Seiberg, \npb{326}{1989}{108};
W. Ogura, \plb{229}{1989}{61}.};
and the connection between $SL(2,R)\simeq SU(1,1)$
Chern-Simons/WZW theory and gravitational Ward identities
in 1+1 dimensions
\nref\polgrav{A.M. Polyakov, \mpl{A2}{1987}{893};
V.G. Knizhnik, A.M. Polyakov, A.B. Zamolodchikov \mpl{A3}{1988}{819}.}%
\nref\verlinde{H. Verlinde, \npb{337}{1990}{652}.}%
\nref\carliou{S. Carlip, \npb{362}{1991}{111}.}%
\refs{\polgrav,\verlinde,\carliou}.  

Because the system has less than maximal supersymmetry,
the supergravity theory has matter fields that are not 
in the same supermultiplet as the graviton.  We will find
in section 5 that the chiral ring of the brane theory
couples to such matter fields, although a complete catalogue
is not generated.
As in \refs{\gkp,\wittenads},
there appears to be a close connection 
between this geometrical sector of the brane theory 
(the chiral ring and its descendant fields) and the 
dimensionally reduced, low-energy supergravity/matter theory.
The many results on absorption into branes
\nref\dasmathur{S.R. Das and S.D. Mathur, hep-th/9606185,
\npb{478}{1996}{561};
A. Dhar, G. Mandal, and S. Wadia, hep-th/9605234,
\plb{388}{1996}{51}.}%
\nref\maldastrom{J. Maldacena and A. Strominger, hep-th/9609026,
\prd{55}{1997}{861}.}%
\nref\maldastromang{J. Maldacena and A. Strominger, hep-th/9702015;
\prd{56}{1997}{4975}.}%
\nref\mathurang{S.D. Mathur, hep-th/9704156.}%
\nref\gubser{S. Gubser, hep-th/9704195; \prd{56}{1997}{4984}.}%
\nref\fxsclrs{C.G. Callan, S.S. Gubser, I.R. Klebanov, and A.A. Tseytlin,
hep-th/9610172, \npb{489}{1997}{65}.;
I.R. Klebanov and M. Krasnitz, hep-th/9612051, \prd{55}{1997}{3250};
I.R. Klebanov and M. Krasnitz, hep-th/9703216, \prd{56}{1997}{2173};
I.R. Klebanov, A. Rajaraman, and A.A. Tseytlin, hep-th/9704112,
\npb{503}{1997}{157}.}%
\refs{\dasmathur-\fxsclrs}
support the idea of a duality between the 
low-energy supergravity/matter theory
and the chiral correlators of the brane theory, since
emission and absorption probabilities
can be calculated either as a chiral correlators on the brane,
or as transmission and reflection amplitudes in the 
supergravity/matter theory; the results are the same.

After assembling the key ingredients of
the relation between CFT data on the brane
and supergravity/matter perturbations in the bulk, 
in section 6 we propose a map between the supergravity theory 
and the brane conformal field theory;
roughly, the partition function of the brane conformal field
theory in an arbitrary background worldsheet metric
and gauge field, with couplings
to chiral operators turned on, is expected to be the wavefunctional
of the corresponding bulk supergravity/matter system,
with the extra dimension of \ads\ spacetime appearing from the
dependence of the CFT partition function on metric data
(the $S^3$ arises from the background gauge field).
This is strikingly similar to the way the matrix model
of noncritical string theory `grows' an extra dimension
(for an extensive review and further references, see
\ref\noncrit{P. Ginsparg and G. Moore, {\it Lectures on
2d gravity and 2d string theory}, hep-th/9304011; in Recent
Directions in Particle Theory (TASI 1992), J. Harvey and J. Polchinski
(eds.), World Scientific (1993).});
indeed, we will find a plethora of parallels between
the two.


\newsec{The D1-D5 system}

The low-energy dynamics of the bound states of $Q_1$ D-strings and 
$Q_5$ D5-branes in type IIB string theory is described 
(taking the four spatial directions
along the D5-branes that are transverse to the D1-branes
to be compactified on a small torus)
by the Higgs branch 
of 1+1 dimensional $SU(Q_1)\times SU(Q_5)$ gauge theory coupled to
hypermultiplets in the $(Q_1,{\bar Q}_5)$ and its conjugate,
as well as adjoint hypermultiplets in each group.
The infrared limit of this system is an $\NN=(4,4)$
sigma model on a target space $\MM$ which is a blowup of the orbifold
$S^{Q_1Q_5} T^4$ \dodf.
The sigma model fields will be denoted
$Y^i_\A$, $\psi_{+\,\A}^{a\alpha}$, $\psi^{\A}_{-\,b\dot\beta}$,
$A=1,...,Q_1Q_5$. 
Here $a,\dot b$ denote spinor indices, and $i$ a vector index, in the
tangent group of the $T^4$; and $\alpha,\dot \beta$ are spinor indices 
of the space transverse to both the worldsheet and the $T^4$.
The space $\MM$ is hyperK\"ahler, hence the IR sigma model is
conformally invariant.  
Further compactification of the spatial coordinate of this
gauge theory on a circle of radius much larger than those
of the internal $T^4$ puts the conformal field theory in finite
volume; then the asymptotic level density for $L_0\gg {\bar L}_0$
computes the black hole entropy
\ref\stromvafa{A. Strominger and C. Vafa, hep-th/9601029;
\plb{379}{1996}{99}.}.
Excitations of the fermions in the sigma model account
for the entropy of spinning black holes
\ref\spinbh{J.C. Breckenridge, D.A. Lowe, R.C. Myers, 
A.W. Peet, A. Strominger, C. Vafa, hep-th/9603078,
\plb{381}{1996}{423};
J.C. Breckenridge, R.C. Myers, A.W. Peet, C. Vafa, hep-th/9602065,
\plb{391}{1997}{93}.}.

An alternative presentation of this 1+1 field theory arises
after T-duality along two circles of the internal $T^4$ (coordinates
$x^6,...,x^9$), 
which turns the D1-D5 system into a set of intersecting
3-branes, $Q_1$ of which are wrapped around (say) the 67
direction and $Q_5$ of which wrap the 89 direction.
The Higgs branch is now realized as the geometrical phase where
the intersection loci are blown up, making a single large
Riemann surface whose genus is $Q_1Q_5$
\ref\cmbi{C.G. Callan and J. Maldacena, hep-th/9708147.}.
The $Q_1Q_5$ hypermultiplets are the blowup modes of the 
degenerate Riemann surface given by the brane construction 
(very similar configurations have been used to study $\NN=2$ 
gauge theories in four dimensions
\ref\wittenneqtwo{E. Witten, hep-th/9703166; \npb{500}{1997}{3}.},
however we are interested in the field theory limit as opposed
to the `MQCD' limit studied there).

The orbifold $S^{Q_1Q_5}(T^4)$ has a large variety of twisted
sectors.  The marginal perturbation of the sigma model on $\MM$
by the $\Z_2$ twist field that interchanges two copies of $T^4$
deforms this manifold to the
desired target space, the moduli space of instantons on $T^4$ \dodf.
While the states of the theory after this perturbation
are no longer eigenstates of the orbifold holonomy,
the predominant configurations intertwine all the copies of $T^4$ 
\ref\maldasuss{J. Maldacena and L. Susskind, hep-th/9604042;
\npb{475}{1996}{679}.}.  
We can thus treat the theory as if it were in
the maximally twisted sector of the unperturbed orbifold.
In this sector, the fields expand
in oscillators having mode numbers $n+\frac{m}{Q_1Q_5}$,
$n,m\in \Z$.  

The chiral ring of this sigma model contains operators
built out of superfields whose lowest components
are the fermions $\psi_{+\A}^{a\alpha}$, $\psi_{-\A}^{a\dot\alpha}$;
these must be contracted with tensor fields
$T^{\sst A_1\cdots}_{a_1\alpha_1\cdots;b_1\beta_1\cdots}$, and
combined into invariants under the symmetric group
acting on the $A$ index.
The superfield whose lowest component is
\eqn\brick{
  \OO^{a\alpha}_{\;b\dot\beta}=
	\sum_{A=1}^{Q_1Q_5}\psi_{+\A}^{a\alpha}\psi^{\A}_{-\, b\dot\beta}
}
generates the chiral ring and its descendants in the untwisted
sector through the operator product expansion.
It also has the nice feature that, when acting on the low-lying states
in the maximally twisted sector, its lowest creation
operators increase the
level number by $1/Q_1Q_5$; in contrast, operators
such as the $SU(2)$ current 
$J_{+}^{\alpha\beta}=\psi_{+\A}^{a\alpha}\psi_{+\A}^{a\beta}$
increase the level number by one.
Thus the `soft' perturbations of the black hole states are
by operators invariant under the orbifold procedure
and of the type \brick, where the $A$ index is
contracted between left- and right-movers.

Highest weight operators preserving an $\NN=2$ subalgebra
of the $\NN=4$ may be constructed as follows.
The $\NN=4$ supercharges are
\eqn\supercurr{\eqalign{
 G_+^{\dot a\alpha}=&~ \sum_{A=1}^{Q_1Q_5}
	\oint \d_+ Y_{b}^{\A\dot a}\psi_{+\A}^{b\alpha}\cr
 G_{-}^{\dot b\dot\beta}=&~ \sum_{A=1}^{Q_1Q_5}
	\oint \d_- Y_{a}^{\A\dot b}\psi_{-\A}^{a\dot\beta}\ .
}}
Denote the components of the SU(2) indices $\alpha,\dot\beta$
by the labels $(1,2)$.
The $l+1$-fold product 
\eqn\hwt{
  \OO^{a_1\cdots a_{l+1}}_{b_1\cdots b_{l+1}}=
	\prod_{i=1}^{l+1} \OO^{a_i1}_{\;b_i1}
}
is annihilated by the supercharges $G_\pm^{\dot a1}$.
Acting by $G_+^{\dot a2}$ and $G_-^{\dot b2}$ for
fixed $\dot a$, $\dot b$ gives the highest component
of a superfield under a particular $\NN=(2,2)$ subalgebra,
which is also highest weight of spin $l$
under the global $SO(4)=SU(2)_L\times SU(2)_R$.
Acting with the global $SO(4)$ lowering
operators $(\oint J_+^{\alpha\beta}\sigma^-_{\alpha\beta})$,
$(\oint J_-^{\dot\alpha\dot\beta}\sigma^-_{\dot\alpha\dot\beta})$
fills out the spin $l$ spherical harmonic.

The simplest chiral operator, 
$\Phi_{\sst (0)}^{ij}=\d_+ Y^i_\A\d_- Y^j_\A$, 
has $l=0$ and was the one used
in the first calculations of absorption by D-brane
black holes
\refs{\dasmathur,\maldastrom}.
The operators
\eqn\phil{
  \Phi_{\sst (l)}^{ij} =
	\gamma^i_{a\dot a}\gamma^{j\,b}_{\ \dot b}
	G_+^{\dot a2} G_-^{\dot b2}
  	\OO^{aa_1\cdots a_{l}}_{ba_1\cdots a_{l}}
}
are similar to those used in
the coupling of the brane theory to higher angular momentum
perturbations in the bulk supergravity 
\refs{\maldastromang-\gubser}.  
There exist many other operators in the chiral ring. 
There are operators with
$h\ne\bar h$, corresponding to 
perturbations with nonzero angular momentum in \ads; 
and with different $SU(2)_L$ and $SU(2)_R$
quantum numbers, needed to couple to higher-spin
fields on $S^3$.  
There are also many operators with higher spin
in the tangent group of $T^4$, obtained by symmetrizing on the
$a,b$ indices of the $\OO^{a\alpha}_{b\dot\beta}$.  
Finally, there is orbifold cohomology
from the twist fields of $S^{Q_1Q_5}(T^4)$.%
\foot{Similar operators
appear in the orbifold construction of 
\ref\kachsilv{S. Kachru and E. Silverstein, hep-th/9802183.};
for instance the operators that split the degeneracy of the 
gauge couplings of the different SU(N) factors in their
construction.}
The simplest such fields come from the
$\Z_2$ twist that interchanges two copies of $T^4$. 
Taking linear combinations $Y_1\pm Y_2$ of the coordinates on
the two $T^4$'s, the odd coordinate 
lives on $T^4/\Z_2$.  In this way one finds
16 basic twist fields which have dimension $h,\bar h=(\hf,\hf)$
and $SU(2)_{L,R}$ spins $j,\bar j=(\hf,\hf)$.  
Because the orbifold is nonabelian, the set of these $\Z_2$ twists
closes on all the higher twist fields (any permutation
can be obtained by a product of transpositions).
Thus, all of the untwisted operators appear in the operator product of
the basic superfield \phil, and all of the twisted operators appear in
the operator product of the basic $\Z_2$ twist superfields;
this fact will be important when we discuss their
role in the AdS/CFT correspondence below.


\newsec{3d supergravity}

The classical geometry of the extremal D1-D5 system compactified
on a four-torus of volume $v$ is
\eqn\cgeom{\eqalign{
  ds^2=&~(H_1H_5)^{-1/2}dx^+ dx^- + (H_1H_5)^{1/2}(d\rr^2+\rr^2d\Omega_3^2)
	+\Bigl(\frac{H_1}{H_5}\Bigr)^{1/2}dx_{{\rm int}}^2\cr
  H_1=&~1+\frac{g\lstr^2 Q_1}{v\,\rr^2}\cr
  H_5=&~1+\frac{g\lstr^2 Q_5}{\rr^2}\ .
}}
The near-horizon geometry is 
$AdS_3\times S^3\times T^4$, with metric
(after a rescaling $r=\rr(\lstr^2g_6\sqrt{Q_1Q_5})^{-1}$)
\eqn\nearhor{
  ds^2=\lstr^2g_6\sqrt{Q_1Q_5}\Bigl[r^2dx^+ dx^- +\Bigl(\frac{dr}{r}\Bigr)^2
	+d\Omega_3^2\Bigr]
	+\bigl(\frac{Q_1}{vQ_5}\bigr)^{1/2}dx_{{\rm int}}^2\ .
}
The radius $\ell$ of the anti-deSitter space and the three-sphere is
$\ell^2=g_6\lstr^2(Q_1Q_5)^{1/2}$, 
whereas the volume of the $T^4$ is $\frac{Q_1}{vQ_5}$; 
$g_6=\gstr\lstr^2v^{-1/2}$ is the effective
fundamental string coupling.
This geometry is semiclassical in the limit $g_6\rightarrow 0$,
with $g_6(Q_1Q_5)^{1/2}$ held fixed and large.%
\foot{It is sometimes said that this geometry yields an exact
string sigma model, since $AdS_3\times S^3\times T^4$
is a group manifold.  However, that sigma model has an NS $B$-field
turned on, whereas the geometry we are interested in has
an RR $B$-field excited.  
Thus the conformal sigma model describes the IR limit of D-string
dynamics near a D1-D5 black string, and does not describe massless
closed strings in the limit of interest $\gstr\rightarrow 0$
(the dual theory with an NS $B$-field would be at
very large coupling). 
Hence the applicability of the group-manifold sigma
model remains to be seen.}
At energies where the Kaluza-Klein modes on $S^3$ are not
appreciably excited, the bulk supergravity theory is 
three-dimensional with a negative cosmological constant.
This theory has 16 supercharges and its infrared
dynamics is conveniently realized
\refs{\hipt,\witcsgrav}
as 2+1 Chern-Simons gauge theory with gauge group 
$SU(1,1|2)\times SU(1,1|2)$; the level is $k={Q_1Q_5}$.
Of the bosonic sector, $SU(1,1)\times SU(1,1)\simeq SL(2,R)\times SL(2,R)$
describes 2+1 anti-deSitter gravity with dreibein $\ee$
and spin connection $\oo$ in terms of `gauge'
fields $A=\oo+\ee/\ell$ and $\tilde A=\oo-\ee/\ell$
\eqn\twodgrav{
  \frac1{16\pi G}\int \ee(\RR-2\Lambda)=\frac{k}{4\pi}\Bigl(
	\int(AdA+\coeff23 A^3)-\int(\atil d\atil+\coeff23\atil^3)\Bigr)
}
up to surface terms.  Here $\Lambda=-\ell^{-2}$, and $k=\ell/(4G)$.
The $SU(2)\times SU(2)$ part of the Chern-Simons gauge group
is the gauged R-symmetry, for which the Chern-Simons term 
is a superpartner of the cosmological constant \hipt\ (in
particular, supersymmetry quantizes the cosmological constant).
There is, of course, the standard kinetic term for these gauge
fields, which is irrelevant in the infrared.
The classical values of the connections $A,\atil$ on \ads\ are
(in the isotropic coordinates \nearhor, in which displacements are 
referred to the \ads\ scale $\ell$) 
\eqn\conn{
  A=\left[\matrix{\hf dr/r& rdx^+\cr
                0& -\hf dr/r}\right] \quad,\qquad
  \atil=\left[\matrix{-\hf dr/r&0\cr
                rdx^-& \hf dr/r}\right] \ .
}

A model of a black hole in \ads\ results from identification of \ads\
under an isometry
\ref\btz{M. Ba\~nados, C. Teitelboim, and J. Zanelli,
\prl{69}{1992}{1849}.  For an extensive review and further
references, see S. Carlip, gr-qc/9506079;
\cqg{12}{1995}{2853}.}.  
One can represent the \ads\ space as a hypersurface in $\IR^{2,2}$ via
\eqn\hyperb{
 \X=\frac1\ell\pmatrix{T_1+X_1&T_2+X_2\cr -T_2+X_2&T_1-X_1}\quad,
	\qquad \det|\X|=1\ .
}
Then the identification $\X\sim \rho_L\X\rho_R$, with
\eqn\idents{
  \rho_L=\pmatrix{e^{\pi(r_+-r_-)}&0\cr 0&e^{-\pi(r_+-r_-)}}
	\quad,\qquad
  \rho_R=\pmatrix{e^{\pi(r_++r_-)}&0\cr 0&e^{-\pi(r_++r_-)}}\quad,
}
has the characteristics of a black hole in \ads, 
with (dimensionless) mass and `angular momentum' 
\eqn\bhparams{
  M={r_+^2+r_-^2}\quad,\qquad
  J=\frac{2\ell r_+r_-}{8G}\ .
}
This geometry appears as the near-horizon geometry of
five-dimensional near-extremal black holes obtained
by compactification of the spatial coordinate $\phi$
($x^\pm=t\pm\phi$), which
have been the subject of extensive study
\refs{\dodf,\maldathesis}.
A set of gauge potentials for this solution is
\eqn\bhconn{
  A=\left[\matrix{\hf d\rho&z_+e^\rho dx^+\cr
                z_+e^{-\rho}dx^+& -\hf d\rho}\right] \quad,\qquad
  \atil=\left[\matrix{-\hf d\rho&z_-e^{-\rho} dx^-\cr
                z_-e^{\rho}dx^-& \hf d\rho}\right] \ ,
}
where $z_\pm=r_+\pm r_-$.  The coordinate $\rho$
covers the exterior of the horizon,
$r^2=r_+^2\ch^2\rho-r_-^2\sh^2\rho\ge r_+^2$.
The $\rho$ coordinate is singular 
in the extremal limit $r_\pm\rightarrow 0$,
in which case it is better to use the coordinate $r$.  Then
the extremal $M=J=0$ background is just \conn, with $\phi$
periodically identified.
Similarly, the extremal $\ell M=8GJ\ne 0$ limit 
is described by the simple modification
\eqn\Jextconn{
  A=\left[\matrix{\hf (dr/r)& rdx^+\cr
                \frac{8GJ}{\ell r}dx^+& -\hf (dr/r)}\right] \quad,\qquad
  \atil=\left[\matrix{-\hf (dr/r)&0\cr
                rdx^- & \hf(dr/r)}\right] \ .
}
The substitution $J\rightarrow p(x^+)$ describes
a family of travelling wave solutions 
\ref\garf{D. Garfinkle and T. Vachaspati, \prd{42}{1990}{1960};
D. Garfinkle, \prd{46}{1992}{4286}.}.

The group of diffeomorphisms that preserves the asymptotic form
of the \ads\ metric \conn\
was shown by Brown and Henneaux to constitute
the Virasoro algebra of conformal transformations of the 
boundary at infinity
\nref\bh{J. Brown and M. Henneaux, \cmp {104}{1986}{207}.}%
\nref\banados{M. Ba\~nados, hep-th/9405171; \prd {52}{1996}{5816}.}%
\nref\strom{A. Strominger, hep-th/9712251.}%
\nref\bbo{M. Ba\~nados, T. Brotz, and M. Ortiz, hep-th/9802076.}%
\refs{\bh-\bbo}, with central charge $c=6k$.
A simple way to see this 
\ref\chvd{O. Coussaert, M. Henneaux, and P. van Driel, 
gr-qc/9506019; \cqg {12}{1995}{2961}.}
is to note
that the 2+1 Chern-Simons theory on a manifold with boundary
is essentially a 1+1 chiral WZW model for the corresponding group \wzw;
furthermore, the asymptotic boundary conditions --
that $A$, $\tilde A$ approach \conn\ as $r\rightarrow\infty$ --
are equivalent \chvd\
to the restriction on the WZW currents that yields the
Hamiltonian reduction of the $SL(2,R)$ WZW model to 
Liouville theory 
\ref\aleks{A. Alekseev and S. Shatashvili, \npb{323}{1989}{719};
M. Bershadsky and H. Ooguri, \cmp{126}{1989}{49}.}.
The Liouville action is the classical generating function
for Virasoro Ward identities.  
Thus one understands the presence of the asymptotic
symmetry group as a consequence of the $SL(2,R)$
WZW/Liouville theory carrying classical
central charge $c=6k$ on the boundary at infinity.
The canonical generators $T_{\pm\pm}$
of the Virasoro symmetry are \refs{\banados,\bbo}
\eqn\cangen{\eqalign{
  T_{++}=&~ -\frac{k}{2}{\rm Tr}\bigl(\alpha^2 + 
	2\alpha\d_+ A_+ + A_+A_+\bigr)\cr
  T_{--}=&~ -\frac{k}{2}{\rm Tr}\bigl(\alpha^2 + 
	2\alpha\d_- \atil_- + \atil_-\atil_-\bigr)\ ,
}}
where $\alpha$ is a constant in the Lie algebra 
such that ${\rm Tr}\alpha^2=\hf$ (\eg\ $\alpha=\hf\sigma^3$).
The supersymmetrization of
these results is straightforward; the group 
$SU(1,1|2)\times SU(1,1|2)$ is the global part of the $\NN=(4,4)$
superconformal group in 1+1 dimensions, so the Liouville theory
generalizes to the $\NN=(4,4)$ Liouville theory
\ref\kpr{C. Kounnas, M. Porrati, and B. Rostand,
\plb{258}{1991}{61}.}.  

The conformal algebra also makes its appearance
in the 2+1 black hole \refs{\banados-\bbo}.
Loosely speaking, the main effect
of the identification \idents\ is to map the conformal
algebra from the conformal plane at the boundary of \ads\
to the conformal cylinder at the boundary of the
black hole spacetime \hyperb,\idents.
This results in a shift of $L_0$ by $h_{\rm min}=c/24$
due to the Schwarzian of the conformal transformation involved.
One has 
\eqn\mass{
  M=\frac{8G}{\ell}(\lz+\lzb)\quad,\qquad J=\lz-\lzb
}
(and thus the `angular momentum' $J$ is the third (`momentum') BPS charge
carried by five dimensional black holes);
equivalently, 
\eqn\rpm{
  r_\pm=\frac{4G}{\ell}\Bigl[\Bigl(\frac{c\lz}{6}\Bigr)^{1/2}
	\pm\Bigl(\frac{c\lzb}{6}\Bigr)^{1/2}\Bigr]\ .
}  
Because of the shift in $\lz,\lzb$, 
anti-deSitter space has `negative mass'
$\lz=-c/24$, or $M=-1$.
The global supersymmetries preserved by \idents\
are periodic in $\phi$
\ref\cousshen{O. Coussaert and M. Henneaux, hep-th/9310194; 
\prl{72}{1994}{183}.};
the black hole has the `Ramond' supersymmetry
appropriate to a cylindrical geometry.

On the basis of the asymptotic symmetry group with its
classical central charge $c=6k$, it has been argued 
\nref\bssent{D. Birmingham, I. Sachs, and S. Sen, hep-th/9801019.}%
\nref\otherent{V. Balasubramanian and F. Larsen, hep-th/9802198;
N. Kaloper, hep-th/9804062.}%
\refs{\strom,\bbo,\bssent,\otherent}
that one can compute the black hole entropy 
\eqn\bhent{
  S=\frac{2\pi r_+\ell}{4G}
}
purely from the algebra of diffeomorphisms, independent of 
considerations of supersymmetry or string theory.
However, having a Virasoro algebra with central charge $c$
implies the asymptotic level density 
\eqn\levdens{
  S\sim 2\pi\Bigl(\frac{c\lz}{6}\Bigr)^{1/2}
   +2\pi\Bigl(\frac{c\lzb}{6}\Bigr)^{1/2}
}
only if the underlying conformal field theory is unitary.
In fact, the prime counterexample is the Liouville model
\ref\kutseib{D. Kutasov and N. Seiberg, \npb{358}{1991}{600}.},
whose density of states grows as \levdens,
but with $c_{\rm eff}=c-h_{\rm min}=1$ appearing in place of $c$.  
Therefore, one cannot conclude from the Virasoro algebra alone
that the microstates are accounted for.
As we will see below, it is the D1-D5 field theory
that provides the underlying unitary dynamics.

Another calculation of the entropy is that of Carlip
\ref\carlip{S. Carlip, gr-qc/9409052; \prd{51}{1995}{632}.},
who analyzes the $SL(2,R)\times SL(2,R)$ WZW model
that results when a boundary is
placed at the black hole horizon.  In a sense,
one is attempting to compute the black hole entropy as
a kind of partition function of the
black hole `horizon degrees of freedom'.
This calculation is puzzling for a number of reasons.  
It proceeds by imposing
$\lz+\lzb=0$ on the effective WZW model on the horizon.
Carlip argues that the zero modes of the current algebra
contribute a large negative amount in the black hole geometry, 
$\lz=j(1-j)/(k+2)$
with $j\sim k^2r_+$,
allowing for a large oscillator excitation at $\lz+\lzb=0$.
The states with $\lz+\lzb=0$ have the same asymptotic level
density as the Bekenstein-Hawking entropy \carlip.
But also because of this negative shift (as well as the
indefinite metric on $SL(2,R)$), many of 
the states being counted do not have positive norm;
the relation to an honest count of microstates in a unitary theory
such as \stromvafa\ is unclear.  
Another difference with the latter computation is that 
the predominant states in
the effective sigma model of \stromvafa\ are spread almost
uniformly over the spatial coordinate $\phi$,
since the effective radius is $k=Q_1Q_5$ times larger
than that seen by bulk supergravity fields \maldasuss.
In contrast, the states counted in \carlip\ have a high
oscillation number in this spatial direction.
Finally, in the present
context we are instructed to extend this computation to supergravity,
where the boundary WZW model has 8 more fermionic and 
3 more bosonic fields for both left- and right-movers; 
a count of the $\lz+\lzb=0$ states in the horizon
$SU(1,1|2)$ WZW model would no longer find the 
correct entropy.


\newsec{Conformal anomaly inflow}

We have a puzzle: how does the supergravity theory
in the asymptotic region of anti-deSitter space ``know''
about the density of states of the brane dynamics?
One possibility is that the branes are indeed located there.
In this section, we suggest a rather different possibility --
the fact that the gravitational action \twodgrav\
is a Chern-Simons term means that it could transport
the conformal anomaly of the branes to distant parts
of spacetime by an anomaly inflow mechanism.

Chern-Simons gauge theory on a manifold $\M$ with boundary takes
the form \wzw
\eqn\csgt{
  S_{\rm CS}[A]=\frac{k}{4\pi}\int_\M (AdA+\coeff23 A^3)
	-\frac{k}{4\pi}\int_{\partial\M}A_uA_v\ .
}
Here we suppose that the boundary of the 3-manifold $\M$ is parametrized
by coordinates $u$, $v$.
The last term is not gauge invariant; rather,
if $A=g^{-1}\abar g+g^{-1}dg$, then one of the components
of $\abar$ is fixed at the boundary (\eg\ $\abar_u$),
and in addition one obtains the gauged chiral WZW model
\eqn\boundwzw{\eqalign{
  S[A]=&~S[\abar]-kS_{\sst WZW}[g,\abar]\cr
  S_{\sst WZW}[g,\abar]=&~\frac1{4\pi}\int_{\partial\M}Tr[
	(g^{-1}\partial_u g+2\abar_u)(g^{-1}\partial_v g)
	+\frac1{12\pi}\int_\M Tr(g^{-1}dg)^3\ .
}}
In the situation at hand, the chiral \sltwo\ WZW model is known
to be equivalent to the generating functional for
Ward identities in 2d gravity 
\refs{\verlinde,\carliou}.  

Let us pause to review these results.
Ordinarily, in quantizing 
Chern-Simons theory one chooses the holomorphic polarization
where $A_u$ are coordinates of the wavefunction; instead, one
may identify the boundary values of the connection $A$
with the 2d spin connection $\omega$ and zweibein $e$
via $A^{0,\pm}=(\omega,e^\pm)$, and work in the 
nonstandard polarization
where $(e^+_u,e^+_v,\omega_u)$ are coordinates.
Then the Gauss Law constraints $\GG^a=0$ on the wave functional are
precisely the gravitational Ward identities for two-dimensional
diffeomorphisms and local Lorentz transformations,
with Virasoro central charge $c=-6k$ at the classical level
\refs{\verlinde,\carliou,\polgrav}.  
The wave functional $\Psi[e_u^+, e_v^+,\omega_u]$ of $SL(2,R)$
Chern-Simons theory in this unorthodox polarization
satisfies \verlinde\foot{Here and below we sometimes use
coordinates $u,v$ in place of $x^\pm$ to avoid confusion between 
worldvolume and tangent space $\pm$ indices.}
\eqn\csgravconst{\eqalign{
  &\GG^a\;\Psi[e_u^+, e_v^+,\omega_u]=0\ ,\cr
  &\GG^+= \frac{ik}{4\pi}(\d_ue_v^+-\d_v e_u^++\omega_ue_v^+)
	-e_u^+\frac\delta{\delta\omega_u}\cr
  &\GG^-=\d_u\frac\delta{\delta e_u^+}-\d_v\frac\delta{\delta e_v^+}
	-\omega_u\frac\delta{\delta e_u^+}
	+i\frac{4\pi}{k}\frac\delta{\delta\omega_u^+}
		\frac\delta{\delta e_v^+}\cr
  &\GG^0=-i\frac{k}{4\pi}\d_v\omega_u+\d_u\frac\delta{\delta_u}
	+e_u^+\frac\delta{\delta e_u^+}
	+e_v^+\frac\delta{\delta e_v^+}\ .
}}
These constraints enforce the $SL(2,R)$ invariance of the
boundary wave functional.  Parametrizing the zweibein as
\eqn\zwei{\eqalign{
  e^+=&~e^\varphi(du+\mu\; dv)\cr
  e^-=&~e^{\bar\varphi}(dv+\bar\mu\; du)\ ,
}}
the constraints \csgravconst\ are solved by the effective
gravitational action
\eqn\graveffact{
  \Psi[\varphi,\mu,\omega]=\exp\Bigl[\frac{ic}{24\pi}\bigl(
	S_O[\varphi,\mu,\omega]+S_L[\varphi,\mu]+S_V[\mu]\bigr)\Bigr]\ ,
}
where $c=-6k$, and
\eqn\esses{\eqalign{
  S_O[\varphi,\mu,\omega]=&~\int dudv\bigl[\hf\mu\omega^2-\omega(
	\d_v\varphi-\mu\d_u\varphi-\d_u\mu)\bigr]\cr
  S_L[\varphi,\mu]=&~\int dudv\bigl[\hf\d_u\varphi\d_v\varphi
	+\mu\bigl(\hf(\d_u\varphi)^2+\d_u^2\varphi)\bigr)\bigr]\cr
  S_V[\mu]=&~\int dudv\;\frac{\d_v F}{\d_u F}\left(\frac{\d_u^3 F}{\d_u F}
	-2\Bigl(\frac{\d_u^2 F}{\d_u F}\Bigr)^2\right)\ .
}}
In the last equation, $\mu$ is related to $F$ implicitly by the
relations $\d_v f=\mu\d_u f$, and $F=f^{-1}$; equivalently,
$S_V$ is defined by the Virasoro Ward identity
\eqn\virward{\eqalign{
  &\VV\;\exp\bigl[\frac{ic}{24\pi}S_V[\mu]\bigr]=0\ ,\cr
  &\VV=(\d_v-\mu\d_u-2(\d_u\mu))\frac\delta{\delta\mu}
	+\frac{ic}{24\pi}\d_u^3\mu\ .
}}
At the special point $\omega=\varphi=0$, the Beltrami parameter $\mu$
is given in terms of $SL(2,R)$ currents as \polgrav
\eqn\belt{
  \mu(u,v)=J^-(v)-2uJ^0(v)+u^2J^+(v)\ ,
}
and the stress tensor $T=\coeff1{k+2}J^aJ^a+\d J^0$
gives straighforwardly $c\sim-6k$ in the semiclassical limit.
This analysis extends to
$\NN=(4,4)$ supergeometry, with
\sltwo\ replaced by $SU(1,1|2)$ (for the $\NN=1$ case, see 
\ref\ezawa{K. Ezawa and A. Ishikawa, hep-th/9612031;
\prd{56}{1997}{2362}. }).
The bulk supergravity effective action is thus the generating functional
of current insertions in the conformal field theory, 
and therefore universal.  

It is important that the theory induced by Chern-Simons dynamics
on the boundary is the chiral WZW model, describing
the gravitational anomalies of only the
left- or right-moving degrees of freedom; otherwise,
we would have two sets of gravitational fields on the boundary.
It has been argued \chvd\ that the two chiralities
formally combine to make the nonchiral gravitational
effective action.  There may be subtleties with zero modes
common to both left and right sectors, however.
We will leave such questions to future work.

The proposal of \refs{\gkp,\wittenads} couples bulk supergravity to 
gauge theory on the boundary via couplings of the form
$\int H\Phi$, where $H$ is a bulk field, and $\Phi$
is a chiral ring operator in the gauge theory.
Witten proposed that it is natural to associate the gauge theory
dynamics with the boundary at infinity in anti-deSitter space.
We shall take a somewhat different path;
in the end, the picture we will develop bears a striking
similarity with the matrix model of noncritical strings \noncrit.
Consider Chern-Simons supergravity on a manifold which is
locally \ads, with an outer boundary at infinity and an inner
boundary along some (timelike) cylinder,
so that the spacetime is topologically an annulus times $\IR$.  
The IR supergravity
dynamics is described by the action \csgt.
Place on this inner boundary the conformal sigma model that
represents the IR dynamics of the D1-D5 system
(the precise sense in which this is meant will be described below).
This sigma model has a conformal anomaly $\hat c=\coeff23 c=4k$.
A conformal reparametrization on the boundary conformal
field theory generates a gravitational anomaly with
coefficient $\hat c$; this anomaly is cancelled by 
`conformal anomaly inflow'
from a combination diffeomorphism and local Lorentz 
transformation in the bulk (matching
the conformal transformation on the boundary), which generates the
gravitational WZW action with coefficient $-\hat c$.
This is a standard anomaly inflow story
\ref\calharv{C. Callan and J. Harvey, \npb{250}{1985}{427}.};
the novelty here is that 2+1 gravity is itself a Chern-Simons theory
to which the anomaly inflow mechanism applies with respect
to diffeomorphisms and local Lorentz transformations.
For the annulus spatial geometry, the Chern-Simons theory is \wzw
\eqn\annu{
  S[A]=S[\abar]-kS^{\rm inner}_{\sst\rm WZW}[g,\abar]
	+kS^{\rm outer}_{\sst\rm WZW}[g,\abar]\ ;
}
one can think of the inner WZW term cancelling the conformal
anomaly of the matter on the branes, while the latter reconstitutes it
on the outer boundary -- in effect, the conformal
anomaly is transported to the outer boundary
even if the degrees of freedom are elsewhere
(or as we shall suggest in a bit, nowhere in particular).
This fact explains the results of \strom. 

The boundary values of the 2+1 connection should be matched
to the background geometry in which the boundary conformal field
theory is written.  Specifically, on the boundary we have
\eqn\bval{
  (A_u^0,A_u^+,A_v^+)=(\omega_u,e_u^+,e_v^+)
	\quad,\qquad
  (\atil_v^0,\atil_v^-,\atil_u^-)=(\omega_v,e_v^-,e_u^-)\ .
}
This has the intriguing effect that, when we evaluate the
gauge potentials $A$, $\atil$ on their classical values \conn\ for
anti-deSitter space,
the location of the inner boundary in \ads\ translates into
the background worldsheet geometry on the static gauge
D1-D5 sigma model.  
The restriction of the inner boundary to surfaces of constant 
$\varphi$ amounts to a kind of minisuperspace approximation.
The sigma model is in the NS sector in order to
match the global (super)isometries.
A similar story applies for the 2+1 black hole geometry.
Again, to match bulk and boundary supersymmetries,
the boundary conformal field theory is in the Ramond sector
(and dominated by the sector of maximal twist \maldasuss,
as discussed in section 2).
In particular, the $M=0$ extremal black hole state is well-approximated
in the CFT by the Ramond sector ground state of the CFT
obtained by acting on the NS vacuum with the Ramond twist
operator of maximal order $k=Q_1Q_5$.

The matching of the bulk supergravity onto the boundary
determines the bulk $SU(1,1|2)$ gauge fields in terms of 
the expectation values of the corresponding currents 
in the conformal field theory.  Consider for example 
the extremal black hole \Jextconn, or rather its
extension to travelling wave solutions.
The classical stress tensor \cangen\ is 
\eqn\gravT{
  k A^+_u A^-_u=p(u)\ ;
} 
this matches that of the conformal field theory 
(using the identifications \bval,\zwei)
\eqn\bdyvar{
  4\pi A_u^+\frac{\delta S_\sigma}{\delta A_v^+}=
	\frac{\delta S_\sigma}{\delta\mu}=
		{T_{uu}(x^+)}
}
provided we identify $p(x^+)=T_{++}(x^+)$.  There is
a similar story for the general black hole \bhconn.
Note that a Legendre transformation 
$A_u^-=\frac{4\pi i}{k}\,\frac{\delta}{\delta A_v^+}$
relates the gravity and brane calculations.  It is important
in order for the matching to work, that the coordinates in the
bulk and boundary theories must be compatible; for instance,
it was shown in 
\ref\hormarolf{G. Horowitz and D. Marolf, hep-th/9605224; 
\prd{55}{1997}{835}.}
that one can find a coordinate transformation 
of the travelling wave metric that locally puts it into
the vacuum \ads\ form \nearhor.%
\foot{Afficionados of Liouville theory will recognize that the quantities
$p(u)=\sigma^2+\d_u\sigma$; $\sigma=\d_u\log(G)$; $G=(\d_uF)^{-1/2}$
employed in \hormarolf\ are the essential characteristics
of the classical solution to Liouville theory, related to
the coordinate transformation $u\rightarrow F(u)$.}
Similarly,
making a conformal transformation $u\rightarrow f(u)$ 
on the boundary CFT produces a term $\frac{c}{12}\{f,u\}$
in the stress tensor $T_{++}$, where $\{f,u\}$ is the
Schwarzian derivative.  Unless the coordinates match properly,
the bulk geometry will not correctly reflect the stress-energy
of the sources on the boundary (pieces involving the
Schwarzian of the coordinate transformation from one to
the other will be missed).  


\newsec{Coupling to matter}

The dimensionally reduced type IIB supergravity theory 
of course consists of more than
$SU(1,1|2)$ Chern-Simons supergravity.
Matter fields in the bulk couple to the 
chiral ring and its descendant fields on the boundary.
The bosonic sector of the D=6 supergravity theory with 32 supercharges
contains the graviton $g_{\mu\nu}$, 5 tensor fields $B_{\mu\nu}$,
16 vectors $A_\mu$, and 25 scalars parametrizing the coset
manifold $\frac{SO(5,5)}{SO(5)\times SO(5)}$.  
The graviton modes $h_{ij}$ on the internal $T^4$
are minimally coupled scalars, and couple
to the $4k$ scalars $Y^i_\A$
($i=1,...,4$; $A=1,...,k$) on the boundary conformal
field theory via the interaction term (for modes that
are S-waves on the $S^3$) {\dasmathur}
\eqn\effint{
  V_{l=0}=\frac1{4\pi\aleff}\int dx^+dx^-\, 
h_{ij}(x^+,x^-)\d_+ Y^i_\A\d_- Y^j_\A\ ,
}
with the effective inverse string tension 
$\aleff=\lstr^4g_6(k)^{1/2}\equiv\ell^2$
\refs{\gubser,\mathurang}; $r_0$ is
the location of the boundary.
(The direct coupling between the internal metric and these
fields is perhaps easiest to see in the geometrical
representation of the Higgs branch 
in terms of D3 branes discussed in section 2.)
Absorption coefficients have been calculated in the \ads\
black hole geometry and compared with this coupling in
\ref\bis{D. Birmingham, I. Sachs, and S. Sen, hep-th/9707188.}.
Replacing $h_{ij}\rightarrow h_{ij}+B^{\sst(RR)}_{ij}$
incorporates the RR $B$-field polarized along $T^4$.

For higher angular momenta, an interaction 
\refs{\maldastromang-\gubser}
has been suggested for angular momenta 
$(\frac l2,\frac l2)\in SU(2)\times SU(2)$ of the form 
\eqn\effintang{
  \frac1{4\pi\aleff}\int dx^+dx^-\,   
	e^{-l\varphi} h^{(l)}_{ij}(x^+,x^-)\Phi^{ij}_{(l)}\ ;
} 
The factor of $exp[-l\varphi]$ is the appropriate
gravitational dressing in conformal coordinates.%
\foot{In arbitrary coordinates, one must replace
$\nabla=\d\rightarrow (1-\mu\bar\mu)^{-1}
e^{-\varphi}(\d+\mu\bar\d)+\omega+\aa$, and 
$\det\, e=e^{2\varphi}\rightarrow (1-\mu\bar\mu)e^{2\varphi}$.
The background gauge field $\aa$ covariantizes
the transformation properties under $SU(2)_{L,R}$;
we will suppress it for now to simplify the discussion.}
The worldsheet operators coupled to $h^{(l)}_{ij}$
are elements of the chiral ring 
given in section 2.  
As an aside, recall that it was also proposed in \gubser\ that, in
the effective string picture, there is
an upper bound $l_{\sst\rm max}=k$
on the $SO(4)$ angular momenta that can be absorbed by the
black string without further suppression by powers of the
frequency $\omega$; however it was unclear what property
of general relativity was responsible for this suppression.
The point is that the higher partial waves are spherical
harmonics on the $S^3$, which couple to the $SU(2)\times SU(2)$
Chern-Simons gauge fields of the effective supergravity
theory (these gauge fields are modes of the graviton in
the original six-dimensional supergravity before Kaluza-Klein
reduction to the IR theory on \ads); the $SU(2)$
representations are only integrable (unitary) for $l\le k/2$.

There are of course many other modes of the reduced supergravity
theory, some of which have been discussed in the literature
\fxsclrs.
The couplings found to date fit the pattern 
of \refs{\gkp,\wittenads} -- they couple
bulk operators to the chiral ring of the 
boundary conformal field theory and its descendant fields.
For example, the $s$-wave mode of the fixed scalar describing
the volume fluctuations of the internal torus
couples to the sum (over $A$) of the product of
energy-momentum tensors $T^\A_{++}T^\A_{--}$;
and the $s$-wave mode of the vector field $V_{\mu i}$
(called an `intermediate scalar' in \fxsclrs), arising from
the components of the metric and $B^{\sst(RR)}$-field
with one index in the internal torus and one index in \ads,
couples to 
$\d_-Y^{\dot b}_{\A a}T_{++}^{\A}$.
The 16 $\Z_2$ twist fields discussed in section 2 have 
the appropriate quantum numbers to be identified with the 16 vector
fields $A_\mu$.
Much of the structure expected from the supergravity side
has been reproduced by coupling the bulk modes to particular
scaling operators on the boundary conformal field theory
\refs{\dasmathur-\fxsclrs}.
It is important to realize that there are vastly many more 
operators in the chiral ring than there are in the supergravity
multiplet.  The latter contains only spins up to two
on the internal $T^4$, whereas very high spin operators
are obtained in the products of the $\OO^{a\alpha}_{b\dot\beta}$
of section 2.  The fact that these operators are formed
in the product of the $\Phi_{(l)}^{ij}$ suggests that
they should be considered as coupling to composite operators
in the supergravity theory.
Similarly, the couplings to higher-order twist fields should be viewed
as composites of those for the basic $\Z_2$ twist operators.

\newsec{A matrix model analogy}

The form \effint,\effintang\ of the interaction of bulk scalars
with boundary chiral fields is not quite what one expects
from supergravity.  Rather, the quantity $e^{-l\varphi}h^{(l)}_{ij}$
is the {\it asymptotic} form of the solution of the scalar wave equation
for the corresponding bulk field $H^{(l)}_{ij}(r,x^+,x^-)$,
provided we identify $r$ with $e^{\varphi}$ 
as suggested by the preceding analysis.
The bulk perturbations $H^{(l)}_{ij}$ are scalar fields on \ads\
obeying the wave equation ($z=1/r$)
\eqn\waveq{
  \Bigl(z\frac{\d}{\d z}z^{-1}\frac{\d}{\d z}+\frac{\d^2}{\d x^+\d x^-}
	-\frac{l(l+2)}{z^2}\Bigr) H^{(l)}_{ij}=0\ ,
}
whose solutions are
\eqn\wavesoln{
  H^{(l)}_{ij}=\vareps_{ij}\, \exp[i(p_+x^+ + p_-x^-)]
	\Bigl(\frac{p}{r}\Bigr)K_{l+1}(p/r)\ ;
}
here $p^2=p_+p_-$, and $\vareps_{ij}$ is a polarization tensor.
Naively, one would have expected the worldsheet
operator $\Phi^{ij}_{(l)}$ to be gravitationally dressed
by the metric $e^{-l\varphi}=r^{-l}$; this only
agrees with \wavesoln\ at asymptotically small values
of $p/r$.\foot{It is interesting to note that the scale
at which these two start to differ is $r\sim p$,
the distance scale that \malda\ associates
with a given energy scale in the brane theory.}

This relation -- between the naive gravitational dressing of
the chiral operators, and the solutions to 
the scalar field wave equation --
is strongly reminiscent of the relation between scaling operators
and macroscopic loops in noncritical string theory \noncrit.%
\foot{The similarities between the CFT/AdS correspondence
and 2d gravity have been noticed by many people,
including the authors of \gkp\ and N. Seiberg,
and probably many others.}
There are indeed many qualitative similarities between
the setup we are proposing -- the partition function
on a 3d annular spacetime -- and the macroscopic loop
amplitudes of two-dimensional gravity \noncrit.
The annulus geometry of noncritical string theory has the
property that, when the proper length of one boundary
shrinks to zero size, the loop operator may be expanded
in a series of local operators (which is equivalent to
setting boundary conditions on the microscopic hole).
For a given scaling operator, the partition function --
considered as a function of the length of the other boundary --
is the wave function of the local operator.

Now think of the outer boundary of \ads\ as a `macroscopic loop' 
shrunk to zero size (the analogue of the
boundary length in 2d gravity is the \ads\ radial coordinate $1/r$).  
The boundary conditions set there
define a set of `microscopic' operators.
On the other hand, think of the inner boundary
as another macroscopic loop (of finite size $1/r$).
Consider the two-dimensional sigma model action 
in a background $\NN=(4,4)$ geometry
\eqn\effact{
  S_{\rm 2d}=  S_\sigma[Y,\psi;\omega,\varphi,\mu,\aa] 
	+ \sum_{l=0}^{k}V_l[Y,\psi;\varphi,\omega,\mu,\aa;h^{(l)}]\ .
}
Here $\aa$ is a background gauge field that covariantizes
the $SU(2)_{L,R}$ dependence (see the footnote after eq. \effintang).
Also, we have included only the minimal scalars on the internal torus;
however, the extension to other chiral fields coupling
to other bulk matter fields is obvious.
From this action we may define the wavefunctional of
a three-dimensional effective theory via
\eqn\corresp{
  \Psi_{\rm eff,3d}[\ee,\oo,\AA_{L,R};H^{(l)}_{ij}]
	=\int\DD Y\DD\psi\; e^{-S_{\rm 2d}}\ .
}
In this map, the metric is determined by the map \bval\
between two-dimensional and three-dimensional geometrical data;
the $SU(2)$ connection $\aa$ determines $\AA_{L\,u}=\aa_u$ and
$\AA_{R\,v}=\aa_v$; and
the matter field $H$ is defined in three dimensions on 
the hypersurface determined by the inner boundary:
\eqn\hfield{
  H^{(l)}=H^{(l)}(r=e^{\varphi(x^+,x^-)},x^+,x^-)
	\quad{\buildrel {r\sim\infty}\over\longrightarrow}\quad
	r^{-l}h^{(l)}_{ij}(x^+,x^-)\ .
}
The results of section 4 show that this map is manifestly
correct when all the chiral couplings $H$ vanish;
then \corresp\ is a wavefunctional for Chern-Simons supergravity.
We see yet another similarity to noncritical string theory
in the generation of another dimension of spacetime 
-- the radial coordinate of \ads\ --
through the scale of the background geometry
(this is rather different from the matrix model
of M-theory \bfss).  
In addition, the background gauge field $\aa$ generates the
three-sphere of $AdS_3\times S^3$ via gauge fields
$\aa_u=g^{-1}\d_u g$, $\aa_v=\d_v h\cdot h^{-1}$; the fields $g$, $h$
compensate the transformation properties of $\Phi_{(l)}^{ij}$
under $SU(2)_{L,R}$ transformations.  In conformal gauge,
the field space $\varphi$, $g$ of $\NN=(4,4)$ Liouville
theory, together with the base space
coordinates $x^\pm$, parametrize $AdS_3\times S^3$;
the Liouville fields determine an embedding of the branes' worldvolume
into spacetime.
Curiously, the analogue of KPZ scaling \polgrav\
is just the opposite here --
whereas the boundary conditions on the macroscopic loop of
noncritical strings matched onto relevant scaling operators
in the UV (large negative $\varphi$), 
in 2+1 gravity the analoguous scaling operators
are irrelevant, and the boundary conditions are set in the IR
(large positive $\varphi$).

The claim of \refs{\malda,\gkp,\wittenads} is that 
the effective action of bulk supergravity/matter
at $k=Q_1Q_5=\infty$ is classical supergravity
coupled to the $\frac{SO(5,5)}{SO(5)\times SO(5)}$ 
Kaluza-Klein scalars $H^{(l)}_{ij}$.  
Consequently, in this limit the log of \corresp\
should be the Hamilton-Jacobi functional of this system.
We regard the classical 2+1 gravity/matter theory 
as the analogue of the tachyon effective field theory
of noncritical strings at cosmological constant $\mu=0$.
The quantity playing the role of $\mu$ in the 2+1 theory
is the coupling $k=\ell/4G$, which is being sent
to infinity to make 2+1 gravity classical.
Pursuing this analogy further, we would say that 2+1 gravity/matter
generates the `bulk S-matrix' through the prescription
of Witten \wittenads, whereas the full brane CFT at
finite $k$ constructs the `wall S-matrix'
(a kind of realization of 't Hooft's `brick wall for black holes'
\ref\thooft{G. 't Hooft, \npb{256}{1985}{727}.}).
In other words, at $\mu=0$ in the matrix model, a perturbation
created at large negative $\varphi$ in the Liouville coordinate 
(the UV region of proper size) never returns as it travels toward larger
values of $\varphi$ (the IR region).
The scattering of noncritical strings 
off the Liouville potential -- the `wall S-matrix' --
is nonperturbative in $\mu$.%
\foot{Or more accurately, the expansion of the S-matrix
in powers of $\mu$ about $\mu=0$ is ill-defined.} 
Similarly, at infinite $k$, an infalling matter perturbation
in \ads\ will not return from beyond the event horizon.
The unitary S-matrix for black holes cannot be detected in
perturbative supergravity (the expansion around $k=\infty$).%
\foot{Also in gravity, the vacuum state (both in \ads\ and
the black hole spacetime) seems to have a `double-sided'
nature -- two asymptotic regions can occur.
It would be fascinating if the effect of turning on
finite $k$ is to throw away the other region,
just as the `Seiberg bound' seems to do in noncritical string theory.}

A number of subtleties plagued the development of noncritical
string theory, that seem unavoidable here as well.  One problem
was operator mixing due to contact terms in situations where
linear combinations of allowed scaling dimensions
could sum to zero
\ref\mss{G. Moore, N. Seiberg, and M. Staudacher, \npb{362}{1991}{665}.}.
Precisely that situation occurs here
as well, and we may expect it to complicate the identification
of bulk supergravity/matter and boundary CFT perturbations.
All of the operators considered in section 2 have integer
scaling dimensions, and we concluded that most of them were
composites of the basic supergravity fields 
$H_{ij}$, $B^{\sst (RR)}_{ij}$, \etc\ -- just as one would expect
of couplings generated by contact interactions.


To summarize, we propose that \effact\ describes bulk supergravity
interacting with a set of internal (black hole) degrees
of freedom at low energies, via a kind of functional integral
transform \corresp.  Correlation functions of the 
two dimensional conformal field theory determine those of gravity,
but not typically vice-versa, except (according
to \refs{\gkp,\wittenads}) at infinite $k=Q_1Q_5$.
The absorption calculations of
\refs{\dasmathur-\fxsclrs,\bis}
confirm that the correspondence is working, at least at
the level of two-point correlations.  Specifically,
the greybody factors for emission and absorption,
are obtained on the one hand
in gravity by solving the wave equation
of an incoming or outgoing scalar field
in the background 3d geometry; and on the other hand,
in the 2d conformal field theory they are generated by
the matrix elements of chiral perturbations
like \effint,\effintang\ in the ensemble of microstates
that are the brane description of a black hole.

Our analysis so far has been closer in spirit to the Lorentzian
signature investigations of \gkp.
How does it compare to the Euclidean analysis of \wittenads\ 
for the computation of correlation functions?  
In the present context,
one should continue to hyperbolic three-space $\IH_3$,
which is formally the vacuum solution to $SL(2,\IC)$
Chern-Simons gravity; the inner boundary becomes the 
two-sphere boundary surrounding a horospherical ball,
tangent to infinity, that has been removed.
The bulk fields that couple to the boundary operators are excited
on this sphere, leading to a perturbation of the CFT.
Similarly, the Euclidean \ads\ black hole 
\ref\carteit{S. Carlip and C. Teitelboim, gr-qc/9405070;
\prd{51}{1995}{622}.}
has the topology of a solid torus -- a disk times $S^1$. 
The connections $A$, $\atil$ are the analytic continuations
of \bhconn, with $t_{\sst \rm Eucl}=it$, $r_-^{\sst\rm Eucl}=-i|r_-|$.
Our prescription involves
the wavefunction on a toroidal boundary surrounding the horizon,
obtained by cutting out a hole containing the origin of the disk.
Holonomies of the Euclidean connection \bhconn\
determine the complex modulus of this torus \carteit;
as we have seen, these holonomies
are directly related to the $\lz,\lzb$ eigenvalues
in the sum over states of the CFT.
The pure supergravity theory codes the current sector correlators
of the conformal field theory, which are guaranteed 
to match properly between CFT and supergravity.
One is thus interested in 
the torus partition function of the CFT
with appropriate operators turned on.
The prescription of \wittenads\ is conjectured to
compute the integrated chiral field correlators 
of the Euclidean CFT on the torus.%
\foot{An obvious generalization of these two Euclidean situations
is to quotient $\IH_3$ by a finitely generated discrete
group that the turns the boundary into the covering space
of a Riemann surface $\Sigma$; it would be interesting
to work out the physical significance of this construction.
A natural candidate is multiple \ads\ Euclidean black holes,
one for each handle.}


\newsec{Final remarks}

It is interesting to contrast the model of low-energy black hole
dynamics presented here with the CGHS model
\ref\cghs{C. Callan, S. Giddings, J. Harvey, A. Strominger,
hep-th/9111056; \prd{45}{1992}{1005}.}.
This 1+1 dimensional model of black hole formation and evaporation
can be cast as the interaction of bulk matter fields
with a dynamical boundary
\ref\mirror{T.-D. Chung and H. Verlinde, hep-th/9311007; 
\npb{418}{1994}{305}.}.
Conservation of stress-energy forces the mirror to accelerate
away from incoming radiation, and above a certain threshold
the mirror accelerates away forever, modelling a black hole
since the incident radiation never returns.
The present model differs from the CGHS model in that
the reflecting wall has many internal
degrees of freedom, which absorb and thermalize the incoming
radiation rather than trying to immediately reflect it back
to infinity.  These internal degrees of freedom are
lacking in the CGHS model (and in general relativity
in higher dimensions). 
These degrees of freedom are also what one needs to account for
the entropy.  Because we have all the relevant low-energy
degrees of freedom, it is hoped that an analysis
of the dynamics of our model will help explain what
fails in the standard field-theoretic analyses of the information problem,
and how string theory describes physics beyond the horizon.
This region would appear in an analytic continuation 
of the background geometric data of the 2d conformal field theory,
thus realizing a speculation of 
\ref\limart{M. Li and E. Martinec,
hep-th/9703211, \cqg{14}{1997}{3187};
hep-th/9704134, \cqg{14}{1997}{3205};
hep-th/9709114.}
as to how matrix theory encodes this part of spacetime.

The analysis of near-extremal five dimensional black strings
completely parallels that given above, so we will not
repeat it.  The brane theory
is an $\NN=(4,0)$ superconformal field theory in the infrared,
with $\hat c=4k=4Q_1Q_2Q_3$ in terms of a triplet of
brane numbers $Q_i$;
correspondingly, the supergravity theory in the bulk
involves the Chern-Simons action for the group 
$SU(1,1|2)\times SL(2,R)$.

Although our considerations have made heavy use of rather special
properties of two dimensional conformal field theory and 
three dimensional gravity, one is led to suppose that
the same thing is happening in the D3 brane 
system with $SU(2,2|4)$ supersymmetry as well.
There appears to be a similar connection between the
scaling properties of the brane theory and the 
Schwinger terms in the algebra of stress tensors
\ref\gubkleb{S.S. Gubser and I.R. Klebanov, hep-th/9708005.}
(see also
\ref\fmmr{D.Z. Freedman, S.D. Mathur, A. Matusis, and L. Rastelli,
hep-th/9804058.}).
The analogue of $e^\varphi$ in our considerations
should be played by the conformal scale factor $f$ of \wittenads.


\vskip 1cm
\noindent{\bf Acknowledgments:} 
The author has benefitted enormously from extensive 
collaborative discussions and correspondence 
with D. Kutasov.  This work would not have come to fruition
without his perceptive comments and suggestions.
The author also thanks 
A. Dabholkar,
S. Das,
J. Harvey,
and
N. Seiberg
for discussions.
This work was supported by DOE grant DE-FG02-90ER-40560.


\vskip 1cm
\noindent{\bf Note added:}
As the manuscript of this article was being finalized,
the author became aware of related work 
by J. Maldacena and A. Strominger, hep-th/9804085; 
and by H. Liu and A.A. Tseytlin, hep-th/9804083.


\listrefs
\end